\documentclass[aps,prd,superscriptaddress,showpacs,floatfix,twocolumn]{revtex4-1}
\usepackage{epsfig}
\usepackage{subfigure}
\usepackage[utf8]{inputenc}
\usepackage{float}
\usepackage{graphicx}
\usepackage{amsmath}
\usepackage{verbatim}
\usepackage{xcolor}
\usepackage{caption}
\usepackage{color}
\usepackage{soul}

\newcommand{\bc}[1]{\textcolor{black}{#1}}

\newcommand{\rd}[1]{\textcolor{black}{#1}}
\begin{document}

\title{Sign Reversal of Boer-Mulders Functions from Semi-inclusive
Deep-Inelastic Scattering to the Drell-Yan Process}

\author{Jen-Chieh Peng}
\affiliation{Department of Physics, University of Illinois at
Urbana-Champaign, Urbana, Illinois 61801, USA}
\affiliation{Department of Physics, National Central University, Taoyuan City, 320317, Taiwan}
\author{Ming-Xiong Liu}
\affiliation{Physics Division, Los Alamos National Laboratory, Los Alamos,
New Mexico 87545, USA}
\author{Guanghua Xu}
\affiliation{Physics Division, Los Alamos National Laboratory, Los Alamos,
New Mexico 87545, USA}
\affiliation{Physical Science Department, South Texas College, McAllen, TX 78501, USA}

\date{\today}

\begin{abstract}
A striking prediction of QCD on the properties of the novel Transverse
Momentum Dependent (TMD) distribution functions is that the time-reversal odd
Sivers and Boer-Mulders functions extracted from semi-inclusive deep-inelastic
scattering (SIDIS) will undergo a sign reversal in the Drell-Yan (DY) process.
This prediction has been tested by experiments that have focused 
on the Sivers functions so far. We \bc{review} the 
current status on the
theoretical prediction and experimental extraction of the signs of
the Boer-Mulders functions from SIDIS and DY. We \bc{note} 
that the existing 
SIDIS and DY data are consistent with the predicted sign reversal of
the Boer-Mulders functions for proton's valence quark distribution. 
Prospects for future experiments 
at EIC capable of testing the sign reversal of the pion Boer-Mulders functions 
are also discussed. 
\end{abstract}


\maketitle

Extensive efforts have been devoted to the study of transverse 
momentum dependent (TMD) parton distributions in 
nucleons during the past decades~\cite{barone02,barone10,bacchetta07}. 
These novel
TMDs are required to describe nucleon's structure functions when quarks
possess non-zero transverse momentum $\vec k_T$, with respect to nucleon's
momentum. Among the various TMDs, the Sivers functions~\cite{sivers90} and the
Boer-Mulders (BM) functions~\cite{boer98} are time-reversal-odd objects and have
attracted much attention both theoretically and experimentally~\cite{perdekamp15,peng14,aidala13}. 

\bc{The Sivers functions represent a correlation between quark's $\vec k_T$
and nucleon's transverse spin~\cite{sivers90}, while the BM
functions signify a correlation between quark's $\vec k_T$
and quark's transverse spin in an unpolarized hadron~\cite{boer98}.
Specifically, the BM function, $h_1^{\perp q} (x,k_T^2)$, is defined 
as~\cite{trento}} 
\begin{equation}
f_{q^\uparrow/h}(x,k_T) = 
\frac{1}{2}[f^q_1(x,k^2_T)-h_1^{\perp q} (x,k_T^2)\frac{(\hat P \times 
\vec k_T) \cdot \vec S_q}{M}],
\label{eq:trento}
\end{equation}
\bc{where $f_{q^\uparrow/h}$ is the distribution of a transversely polarized
quark in an unpolarized hadron $h$ and $f^q_1$ is the unpolarized quark
density. Here $\hat P$ is the momentum vector of the hadron, $\vec S_q$
the spin vector of the quark, and $M$ the mass of the hadron. The sign
of the BM function follows the Trento convention~\cite{trento}.}

Although the very existence of the time-reversal-odd TMDs
was in question at one time~\cite{collins93},
it was later shown that these functions can arise from initial- or
final-state interactions~\cite{brodsky02}.
Such interactions are incorporated in a natural fashion by gauge
links that are required for a gauge-invariant definition of 
TMDs~\cite{collins02,belitsky03}.
Measurements of the semi-inclusive
deep-inelastic scattering (SIDIS) at 
HERMES~\cite{hermes05,hermes09,hermes10},
COMPASS~\cite{compass09,compass12,compass15,compass24a} and 
JLab~\cite{xin11,zhao14} using transversely polarized targets
have shown clear evidence for the presence of the T-odd Sivers
functions. These data also allow the
extraction~\cite{anselmino12,anselmino13,martin15,kang16,anselmino15,ethier17,
martin17} of the momentum and
flavor dependencies of the Sivers functions.

The gauge-link operator leads to a remarkable prediction~\cite{collins02}
that the signs of the 
T-odd Sivers and BM functions are process dependent, namely, 
they must have opposite signs depending on whether they are involved
in the space-like SIDIS or the time-like
Drell-Yan (DY) process~\cite{peng14}. An experimental verification of the sign-reversal
prediction of the Sivers and BM functions would provide an important
test of QCD at the confinement scale and represents a significant step 
towards understanding the properties of these novel TMDs.

Several DY experiments have been proposed or performed
to test the predicted sign reversal of the Sivers functions.
At RHIC, transversely polarized proton beams
allow measurements of transverse
single-spin asymmetries (TSAs) in $p + p$ DY,
$W$-boson, and $Z$-boson productions. The first results on TSAs
of $W$ and $Z$ production at RHIC energy have been reported
by the STAR Collaboration~\cite{star16,star24}.
The COMPASS experiment has
measured the TSAs for the DY process using
a 190 GeV/c $\pi^-$ beam on a transversely polarized NH$_3$  
target~\cite{compass17,compass24}.
The SpinQuest experiment at Fermilab plans to use a 120 GeV
proton beam to measure the TSAs for the DY process
on a transversely polarized NH$_3$ target~\cite{klein13}.
Within the experimental uncertainties, both the STAR and the COMPASS
measurements are consistent with
the QCD prediction of a sign reversal of the Sivers functions. However, a conclusive test for the sign reversal of the Sivers function awaits
future experiments~\cite{aidala19,barschel20}.  

Although the subject of the sign reversal of 
the Sivers functions has been
discussed extensively in the literature, 
relatively little attention has been paid to the possibility of
testing the sign reversal of the BM functions.
This probably reflects the fact that the BM functions have
not yet been well determined from existing SIDIS data.
Nevertheless, we note that there already exists some information on
the BM functions from the unpolarized
DY experiment~\cite{boer99}. 
Indeed, BM functions were the first TMD  
measured in the DY experiments~\cite{ma_pion,ma_sea}.
It is important to understand how existing
SIDIS and DY data could test 
the predicted sign reversal of the BM functions.
In this paper, we show that 
the existing data are in favor of the predicted
sign reversal of the proton BM functions. We 
also suggest possible measurements at the future
Electron Ion Collider (EIC) 
to test the predicted sign reversal of the
pion BM functions. 

We first briefly review the theoretical expectations 
on the signs and quark-flavor dependence of the BM functions
for nucleons and pions. We then examine the current status 
of the determination of the sign and magnitude of the BM functions 
from the SIDIS and DY experiments. The prospects
for testing the sign-reversal prediction of the BM functions
will then be presented.

Unlike the
parton density distributions, which are positive-definite, the TMDs
can have positive or negative signs. Using the sign convention
in~\cite{trento} for the TMDs, the Sivers functions
for proton's valence $u$ and $d$ quarks were predicted in many theoretical
models to have opposite signs, namely, negative
for $u$ and positive for $d$, in qualitative agreement with the results
obtained in SIDIS~\cite{hermes09,compass12,compass15}. 
For the nucleon's BM functions, calculations using 
the bag model~\cite{bm-bag-model}, the 
quark-spectator-diquark model~\cite{bm-diquark}, the large-$N_c$ 
model~\cite{bm-nc}, the relativistic constituent quark 
model~\cite{bm-rel}, as well as lattice QCD~\cite{bm-lattice}, all
predict negative signs for
both the $u$ and $d$ valence quarks in SIDIS. 

The Sivers functions, signifying the correlation between the hadron's 
transverse spin direction and the
quark's transverse momentum direction, $\hat k_T$, must vanish for
spin-zero hadrons such as pions and kaons. On the other
hand, the BM functions, being independent of hadron's spin,
can exist for pions and kaons. Calculations for pion's 
valence-quark BM functions 
using the quark-spectator-antiquark model~\cite{ma_pion} and the 
light-front constituent approach~\cite{pasquini14} both  
predict a negative sign,
just like the $u$ and $d$ valence-quark BM functions of the nucleons. Using
the bag model, the valence-quark BM functions for mesons and nucleons 
were predicted~\cite{burkardt08} to have similar magnitude with the same
signs. Since the nucleon's valence-quark BM functions are
predicted to be negative, this implies that pion's valence-quark
BM functions are also negative. This prediction~\cite{burkardt08} 
of a universal behavior of the BM functions for pions and nucleons awaits
experimental confirmation.

For nucleon's antiquark BM functions, there
exists only one model calculation so far. It was pointed out~\cite{ma_sea}
that the nucleon meson cloud could contribute to its sea-quark BM functions.
The clear evidence for the meson cloud as an important source of sea quarks in 
nucleons was provided by the large $\bar d / \bar u$ flavor 
asymmetry observed in DIS and DY experiments~\cite{garvey}.
A significant fraction of nucleon's antiquark sea at the
valence region comes from the meson cloud. This suggests
that pion cloud can contribute to nucleon's
antiquark BM functions~\cite{ma_sea}. The implication is
that nucleon's antiquark BM functions would have negative signs, just
like pion's valence-quark BM functions.

\begin{table}[tbp]   
\caption {Theoretical predictions for the signs of proton's ($p$)
valence and antiquark BM functions in SIDIS 
and Drell-Yan. The prediction for the signs of 
pion's  valence-quark BM
function ($V_\pi$) is also shown.}
\label{tab:sign}
\begin{center}
\begin{tabular}{|c|c|c|c|c|c|}
\hline
\hline
 & ~$u_p$~ & ~$d_p$~ & ~$\bar u_p$~ & ~$\bar d_p$~& ~$V_\pi$~ \\
\hline
\hline
SIDIS & $-$ & $-$ & $-$ & $-$ & $-$ \\
\hline
~~Drell-Yan~~ & + & + & + & + & + \\
\hline
\hline
\end{tabular}
\end{center}
\end{table}

Table~\ref{tab:sign} summarizes the theoretical expectations for
the signs of the BM functions discussed above. First, the valence 
$u$ and $d$ BM functions of the nucleons have negative signs. Second, the 
valence-quark BM functions of the pions are also 
predicted to be negative, just like those of the 
nucleons. Third, the antiquark BM
functions in the nucleons are also negative, based on the
meson-cloud model. Finally, Table I shows that the signs 
of these BM functions will reverse and become positive
for the DY process. 

We now compare the predictions shown in Table I with the experimental results. The BM functions of the nucleons can be extracted from the azimuthal angular
distribution of charged pions produced in unpolarized SIDIS~\cite{boer98}.
At leading twist, the $\cos 2\phi$ term in the angular distribution is proportional to the product
of the nucleon's BM functions $h^\perp_1$ and the Collins fragmentation 
functions $H^\perp_1$ for quarks hadronizing into charged pions. The 
angle $\phi$ refers to the azimuthal angle 
of the produced pion with respect to the lepton scattering plane.
At the low $p_T$ region, the 
$\langle \cos 2 \phi \rangle$ moment has been measured by the 
HERMES~\cite{hermes13} and 
COMPASS~\cite{adolph14}
collaborations. \bc{The average $Q^2$ for the HERMES SIDIS measurement
is 2.4 GeV$^2$ and a somewhat higher value of 3.8 GeV$^2$ for COMPASS.}
\bc{An analysis of these $\langle \cos 2 \phi \rangle$
data for pion SIDIS was performed~\cite{barone10a} by including the
leading-twist contribution of the BM function and the twist-4 contribution
of the Cahn effect~\cite{cahn}. For the low-$P_T$ region, $P_T \le 1$ GeV,
the perturbative QCD effect is expected to be negligible~\cite{chay}.}
The BM functions are assumed to have the following functional 
form~\cite{barone10a}:

\begin{equation}
h_1^{\perp q} (x,k_T^2) = \lambda_q f_{1T}^{\perp q} (x, k_T^2),
\label{eq:bmfit}
\end{equation}
\noindent where $q$ refers to the quark flavor and $h_1^{\perp q}$ and
$f_{1T}^{\perp q}$ are the BM and Sivers functions, respectively. 
Equation~(\ref{eq:bmfit}) 
assumes the same $x$ and $k_T^2$ dependencies for the BM and
Sivers functions with the sign and magnitude of the 
proportionality factor
$\lambda_q$ determined from the data. 
\bc{Equation~(\ref{eq:bmfit}) also
assumes that the BM function has the same $Q^2$ evolution as the
Sivers function, which was assumed to have the same evolution as the
unpolarized parton distribution function $f^q_1$.}
The Sivers functions determined from a fit~\cite{anselmino09a} to 
the polarized SIDIS data together with the Collins 
fragmentation functions from~\cite{anselmino09b}
were used. The analysis~\cite{barone10a}	 
yielded the best-fit values of 
$\lambda_u = 2.0$
and $\lambda_d = -1.1$. Since the Sivers function for $u (d)$ 
is negative (positive), these best-fit values imply that the BM functions 
$h_1^{\perp u}$ and
$h_1^{\perp d}$ are both negative, in agreement with the theoretical 
expectation shown in Table~\ref{tab:sign}. 
\rd{It should be cautioned that the extraction of the BM function in SIDIS
depends sensitively on the treatment of the Cahn effect. Future SIDIS data
taken at higher $Q^2$ to minimize the high-twist Cahn effect is of great
interest~\cite{barone10a}.}

It should be noted
that the signs of the Collins fragmentation functions are not determined
experimentally, since only the product of two Collins
fragmentation functions is measured in the $e^+ e^-$ experiments
at Belle~\cite{ralf} and Babar~\cite{lees14}. The signs of the 
Collins fragmentation functions
were determined in~\cite{anselmino09b} such that the extracted $u$ and
$d$ transversity distributions have the same signs as the 
corresponding $u$ and $d$ helicity distributions (that is, positive 
for the $u$ quark and negative for the $d$ quark transversity distributions).
It is reassuring that the
signs of the extracted $u$ and $d$ BM functions from 
SIDIS~\cite{barone10a} agree with theoretical 
expectation~\cite{bm-bag-model,bm-diquark,bm-nc,bm-rel,bm-lattice}.

\bc{The  best-fit values of
$\lambda_u = 2.0$
and $\lambda_d = -1.1$ in Eq.~(\ref{eq:bmfit}) are consistent with the
theoretical expectation~\cite{bm-lattice,burkardt05} 
that $h^{\perp u}_1$ is larger in magnitude
than the corresponding $u$-quark Sivers function $f^{\perp u}_{1T}$,
and that $h^{\perp d}_1$ and $f^{\perp d}_{1T}$ have comparable magnitudes. 
The $x$-dependence of $x|h^{\perp u,d}_1(x)|$ obtained from the best-fit 
to the SIDIS data, as shown in Fig. 3 of~\cite{barone10a}, peaks at 
$x \sim 0.2$ with values of $\sim 0.08$ and $\sim 0.06$ for $u$ and
$d$ respectively.}

The SIDIS data are not yet able to constrain the proton antiquark 
BM functions, whose contributions are expected to be 
overshadowed by their
quark counterparts. In the analysis of~\cite{barone10a}, 
it was assumed that the antiquark BM functions
were equal in magnitude to the corresponding
Sivers functions with a negative sign, namely,
\begin{equation}
h_1^{\perp \bar q} (x,k_T^2) = -|f_{1T}^{\perp \bar q} (x, k_T^2)|.
\label{eq:bmfita}
\end{equation}
\noindent This ad-hoc assumption would add to the systematic uncertainty
for the analysis of~\cite{barone10a}. Nevertheless, the results 
on the valence-quark BM functions are expected to be largely insensitive 
to this assumption about the antiquark BM functions.

The HERMES collaboration has reported results~\cite{hermes13} on the
azimuthal $\cos 2 \phi$ modulations for $\pi^{\pm}$, $K^{\pm}$,
and unidentified hadrons in unpolarized $e+p$ and $e+d$ SIDIS.
The $K^{\pm}$ and unidentified hadron data were not included in the
previous work~\cite{barone10a} to extract nucleon BM functions.
These new HERMES data could lead to a more precise extraction of 
valence-quark BM functions. In addition, these data are sensitive to
antiquark BM functions. In particular, the $\cos 2 \phi$
moments for $K^-$ production are observed to be large 
and negative~\cite{hermes13}. Since the valence-quark content of $K^-$,
$s \bar u$, is distinct from that of target nucleons, the large
negative $\cos 2 \phi$ moment for $K^-$ suggests sizable 
sea-quark BM functions.
An extension of the global fit in Ref.~\cite{barone10a} to include the 
new $K^\pm$ data would be very valuable and could allow the extraction
of the proton antiquark BM functions in SIDIS. 
Table II summarizes the current experimental knowledge on the signs 
 of proton valence and antiquark BM functions from the
SIDIS.

In order to test the prediction of sign reversal from SIDIS to 
DY for the BM functions, 
we turn next to the extraction of the BM functions
from the DY experiment. BM functions can be 
extracted~\cite{boer99}
from the DY process using an unpolarized or a singly
polarized hadron-hadron collision. 
The expression for the unpolarized DY
angular distribution is~\cite{lam78} 
\begin{equation}
\frac {d\sigma} {d\Omega} \propto 1+\lambda \cos^2\theta +\mu \sin2\theta
\cos \phi + \frac {\nu}{2} \sin^2\theta \cos 2\phi,
\label{eq:eq1}
\end{equation}
\noindent where $\theta$ and $\phi$ are the polar and azimuthal angles
of $l^+$ in the dilepton rest frame. Boer showed~\cite{boer99} 
that the $\cos 2\phi$
term is proportional to the convolution of the quark and antiquark
BM functions in the projectile and target, namely,
\begin{equation}
\langle \cos(2\phi) \rangle \sim \sum_{q,\bar q} [h^{\perp q}_1 (x_1)
h^{\perp \bar q}_1 (x_2) + h^{\perp \bar q}_1 (x_1)
h^{\perp q}_1 (x_2)],
\label{eq:eq2}
\end{equation}
\noindent where $x_1$ and $x_2$ refer to the momentum fractions
of the projectile and target hadrons carried by the partons, 
and the sum is over 
quark flavors. The BM functions for quarks and antiquarks
are denoted as $h^{\perp q}_1$ and $h^{\perp \bar q}_1$.

Pronounced $\cos 2 \phi$ dependencies
were observed in the NA10~\cite{falciano86,guanziroli88} and 
E615~\cite{conway89}
$\pi^-$-induced DY experiments on tungsten and deuterium targets. 
The coefficient $\nu$ for the 
$\cos 2\phi$ term in Eq.~(\ref{eq:eq1}) was found to be positive
with the mean value $\langle \nu \rangle = 0.091 \pm 0.009$ at
194 GeV/c~\cite{guanziroli88} and $0.169 \pm 0.019$ at
252 GeV/c~\cite{conway89} over the $0<p_T<3$ GeV/c range. 
In all $\pi^-$-induced DY 
experiments, $\nu$ was found to be positive. 
Together with Eq.~(\ref{eq:eq2}) and the dominance of
$u - \bar u$ annihilation in the $\pi^-$-nucleus DY
process, the positive sign of $\nu$ suggests two possibilities:
either the signs of pion's and proton's valence-quark BM functions
are both positive, or both negative. Table I shows 
that the first possibility of positive signs agrees with the 
prediction of sign reversal for BM functions in the DY process. 
However, the second possibility, where both pion and proton 
valence-quark BM functions have a negative sign in the DY process,
is consistent with the scenario of no sign reversal.
As discussed later, additional data
are required to distinguish these two possibilities.

The $\cos 2 \phi$ dependencies were also measured in the
$p+p$ and $p+d$ unpolarized DY experiment~\cite{zhu07,zhu09}
\bc{using an 800 GeV proton beam with a mean $Q^2$ of 54 GeV$^2$.}
The magnitude of $\nu$ was found to be significantly smaller than that in the pion DY experiment. Since proton-induced
DY involves both the valence and the sea quarks in the beam and 
target hadrons, the value of $\nu$ now involves the convolution of
the valence-quark BM function \bc{in the beam proton and the sea-quark BM 
function in the target nucleon}. The small values for $\nu$ reflect the subdominance
of sea-quark BM functions and are consistent with the theoretical 
expectation~\cite{ma_sea}.
The signs of $\nu$ for both $p+p$ and $p+d$ DY are found
to be positive~\cite{zhu07,zhu09}. From Eq.~(\ref{eq:eq2}), this
suggests that the proton's sea-quark BM function has the same sign as the valence-quark
BM function, consistent with the prediction shown in
Table I. However, the data could not determine whether the signs are
positive or negative.

\begin{table}[tbp]   
\caption {Experimental information on the signs of proton's valence and
antiquark BM functions in SIDIS and unpolarized DY. 
$V_\pi$ signifies the valence
quarks in the pions. The two separate rows, (a) and (b) for the Drell-Yan,
correspond to two allowed solutions from the existing unpolarized 
data. The polarized DY data from COMPASS~\cite{compass24} 
now favors solution (a).}
\label{tab:sign_ex}
\begin{center}
\begin{tabular}{|c|c|c|c|c|c|}
\hline
\hline
	& ~$u_p$~ & ~$d_p$~ & ~$\bar u_p$~ & ~$\bar d_p$~ & ~$V_\pi$~ \\
\hline
\hline
SIDIS & $-$ & $-$ & no data & no data & no data \\
\hline
	~~Drell-Yan (a)~~ & + & + & + & + & + \\
\hline
        ~~Drell-Yan (b)~~ & $-$ & $-$ & $-$ & $-$ & $-$  \\
\hline
\hline
\end{tabular}
\end{center}
\end{table}

\bc{The antiquark BM functions of the proton have been extracted from an
analysis~\cite{barone10b} of the $p+p$ and $p+d$ DY data. The shape of the 
antiquark BM function is assumed to be the same as the antiquark Sivers
function extracted in~\cite{anselmino09a}, and the valence quark BM function 
are taken to be identical to the valence BM functions obtained in the 
analysis~\cite{barone10a} of the SIDIS data. This analysis shows that the
valence and antiquark BM functions have the same sign. It also shows
that the magnitudes 
of the antiquark BM functions are much smaller than that of the valence
quarks. As shown in Fig. 3 of~\cite{barone10b}, the peak values of 
$x|h^{\perp \bar u, \bar d}_1(x)|$ are $\sim 0.0023 $ and
$\sim 0.0014$ for $\bar u$ and $\bar d$, respectively. These results are in 
good agreement with another independent analysis~\cite{lu10}.}

The current status on the signs of the BM functions
deduced from the SIDIS and DY experiments is summarized 
in Table II. 
A comparison between the predictions listed in Table I
and the experimental status presented in Table II shows that the
data are consistent with theoretical expectations with no
disagreement found. Unfortunately, the inability for the 
unpolarized DY data on $\nu$ alone to distinguish the two
possible solutions on the signs of the nucleon BM functions
prevents the determination of the signs of proton's BM functions in
the DY process. In order to test the prediction of sign reversal for the
BM functions, the key measurements would involve a singly polarized
DY where a nucleon is transversely polarized, as first proposed 
in~\cite{peng15}, and further discussed below.
\rd{We note that evidences for sign reversal of the BM functions in DY, 
based on~\cite{peng15} and the recent polarized DY 
measurements~\cite{compass17,compass24} at COMPASS,
have also been reported in~\cite{compass24,bastami21}.}

We consider the $\pi^-$-induced Drell-Yan process on a
transversely polarized proton target. This measurement was recently
pursued by the COMPASS experiment at CERN~\cite{compass17,compass24} with
the primary goal of testing the sign reversal of the Sivers function.
The DY cross section for pion interacting with a transversely polarized proton target can be written 
as~\cite{metz,sissakian}
\begin{eqnarray}
\frac {d\sigma} {dq^4 d\Omega} & \propto & 1 + S_T 
\left[ D_1
A^{\sin\phi_S}_T \sin\phi_S \right] \nonumber \\
& + & S_T \left[ D_2 A^{\sin(2\phi-\phi_S)}_T\sin(2\phi-\phi_S) 
\right] \nonumber \\
& + & S_T \left[ D_2 A^{\sin(2\phi+\phi_S)}_T \sin(2\phi+\phi_S)
\right],
\label{eq:eq3}
\end{eqnarray}
where $S_T$ is the proton's spin component transverse to
the hadron plane, formed by the momentum vectors of
the beam and target hadrons in the dilepton's
rest frame. $\phi_S$ and $\phi$ refer to the azimuthal angles of the target
spin direction and the charged lepton momentum direction, respectively. The 
amplitudes of various azimuthal angular modulations are indicated
by $A^{m(\phi_S,\phi)}_T$ with $m(\phi_S,\phi)$ specifying the
form of the azimuthal angular modulation. 
$D_1$ and $D_2$ are the depolarization factors.

Equation~(\ref{eq:eq3}) shows that the three amplitudes,
$A^{\sin\phi_S}_T$, $A^{\sin(2\phi-\phi_S)}_T,$ and 
$A^{\sin(2\phi+\phi_S)}_T$, depend on the transverse spin 
direction of the target nucleon. The first amplitude,
$A^{\sin\phi_S}_T$, is a convolution of the nucleon Sivers function
and the unpolarized distribution of the pion. Since pion's unpolarzied parton
distributions are positive-definite, the sign of $A^{\sin\phi_S}_T$
directly reflects the sign of the nucleon Sivers function, allowing a test of the sign-reversal prediction for nucleon Sivers functions.

The other two amplitudes in Eq.~(\ref{eq:eq3}), $A^{\sin(2\phi-\phi_S)}_T$ and
$A^{\sin(2\phi+\phi_S)}_T$, are related to the convolution of the
pion BM function and nucleon's transversity ($h_1$) and
pretzelosity ($h^{\perp}_{1T}$) distributions, respectively. 
For the $\pi^-$-induced DY process on a transversely polarized
proton target, such as in the COMPASS experiment, $u$-quark
dominance implies that $A^{\sin(2\phi-\phi_S)}_T$ is proportional
to the product of the pion's $\bar u$ valence-quark BM function and 
the proton's $u$-quark transversity distribution. Since the sign of 
the proton's $u$-quark transversity distribution is found to
be positive~\cite{Mauro_13}, a measurement of the sign of 
$A^{\sin(2\phi-\phi_S)}_T$ in polarized $\pi^ - p$ DY
would determine the sign of pion's valence-quark BM function.
\bc{Once the sign of the pion's valence-quark 
BM function is known, Table II shows that the sign of proton's $u$-quark 
BM function in the DY process can be determined.} More specifically,
if $A^{\sin(2\phi-\phi_S)}_T$ in polarized $\pi^ - p$ DY
is found to be positive, then the sign of proton's $u$-quark BM function
in the DY process will be positive and the predicted 
sign reversal of BM function will be confirmed. In contrast, 
a negative $A^{\sin(2\phi-\phi_S)}_T$
would cast doubt on the prediction of sign reversal.
\begin{figure}
\includegraphics*[width=\linewidth]{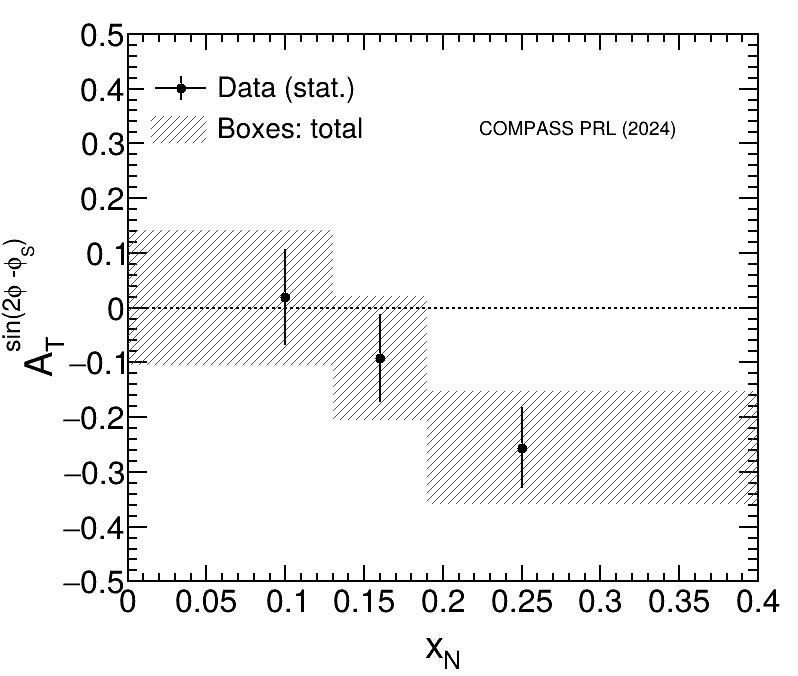}
\caption{The COMPASS measurement for the $A^{\sin(2\phi-\phi_S)}_T$ asymmetry
versus $x_N$ in $\pi^-$-induced DY process
        on transversely polarized proton target~\cite{compass24}, where
	$x_N$ is the momentum fraction carried by the target parton. \bc{The 
	event-weighted average of $x_N$ is plotted for each bin. The horizontal
	width of the boxes indicate the range in $x_N$.}
        }
\end{figure}

\begin{figure}
\includegraphics*[width=\linewidth]{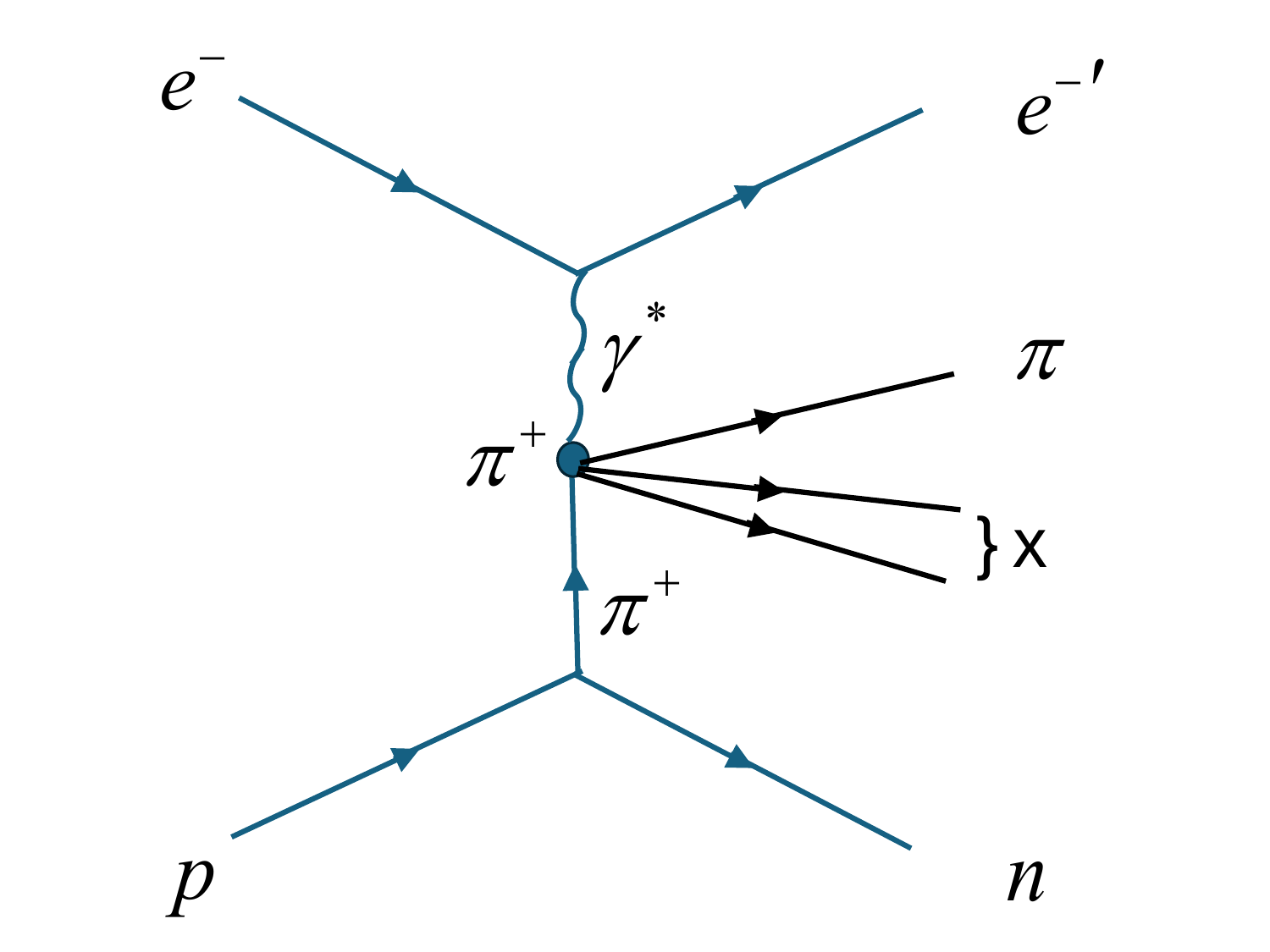}
\caption{Illustration of the SIDIS process on the meson cloud
        via the Sullivan process. The virtual photon emitted from the
        electron would undergo a semi-inclusive process on the pion leading
        to a production of a pion, which is detected, togther with the tagged
        neutron along the beam proton direction.}
 \label{SIDIS_Sullivan}
\end{figure}
The COMPASS results on the azimuthal asymmetry of $\pi^-$-induced DY on a transversely polarized proton target have shown that the 
sign for this $A^{\sin(2\phi-\phi_S)}_T$ is 
negative, with an average value of $-0.131 \pm$ 0.046 (stat) $\pm$ 0.047 (syst)~\cite{compass17,compass24}, as shown in Fig. 1. 
This seems to suggest that there
is no sign reversal for the proton's BM function in the DY process. 
However, the coordinate system adopted by
COMPASS is opposite to the usual convention, namely,
the $z$-axis is along the unpolarized pion beam direction, rather than the
direction of the polarized target nucleon (in the CM frame). Therefore,
the negative sign reported by the COMPASS actually corresponds to a
positive sign in the usual convention. Hence, one can conclude that the
result from COMPASS supports the expectation that the BM
function indeed changes sign in the DY process. \bc{The 
two-fold ambiguity labelled (a) and (b) in Table II is now resolved
with the solution (a) favored over solution (b), based on the polarized
DY measurement at COMPASS.}

We mention in passing that the amplitude $A^{\sin(2\phi+\phi_S)}_T$
in Eq.~(\ref{eq:eq3}), though interesting, would not lead to a determination
of pion's BM function, since the nucleon's prezelocity distribution is yet unknown.

As the latest result from COMPASS suggests that the valence 
BM functions for both
proton and pion have a positive sign in the DY process, it would be 
interesting to check whether
the pion valence-quark BM function also undergoes a sign reversal from SIDIS to DY.
At first sight, it seems impossible to measure the BM function of pion in SIDIS, since pion is not available as a target. Nevertheless, it is conceivable that one could use the Sullivan process to perform SIDIS
on the virtual pion target at the EIC. This would determine the sign of the
pion's BM function from SIDIS, similar to the determination of the sign
of proton's BM function from SIDIS on proton target. Fig. 2 illustrates the SIDIS process on the pion cloud through the Sullivan process at the EIC. The measurement would involve the tagging of the neutron, together with the detection of the scattered electron and the pion produced in the SIDIS on the virtual pion. There have been proposals to measure the PDFs of pion and kaon at the future EIC and the Electron-Ion Collider in China (EicC) using the inclusive DIS reaction on
the meson cloud via the Sullivan process~\cite{EIC-YR, EICC}. An extension of such a DIS measurement to SIDIS measurement could lead to the determination of the sign of pion BM functions. As a spin-zero hadron, the Sivers function vanishes for pion, and the BM function is the only quantity available to test the sign-reversal prediction for the meson sector. 

In summary, in this paper we emphasize the importance of extending the experimental tests of the QCD prediction of sign reversal from the measurement of the 
Sivers functions to the Boer-Mulders functions. There is tantalizing 
evidence from the recent COMPASS data that such a sign reversal indeed 
occurs for the proton valence-quark BM function. A global analysis of 
existing SIDIS and DY data is required before a definitive conclusion 
can be reached. We also point out a possible test of the sign-reversal 
prediction of the pion valence-quark BM function via the Sullivan process 
at the future EIC.

The authors thank Francesca Giodano for her interest during the initial 
stage of this work and Bakur Parsamyan for providing the COMPASS data.
This research is supported by NSF Grant No. PHY 2110229 and DOE Office of 
Science Nuclear Physics Program. G. Xu was supported in part by the U.S. 
Department of Energy, Office of Science, Office of Workforce Development 
for Teachers and Scientists (WDTS) under the Visiting Faculty 
Program (VFP). J. Peng was supported in part by the Yu-Shan program of 
the Ministry of Education in Taiwan.

\end{document}